\documentclass[twocolumn]{aastex63} 
\usepackage{epsfig}
\usepackage{graphicx}
\usepackage{float}
\usepackage{amsmath}
\usepackage{color}
\usepackage{amssymb}
\usepackage{amsfonts}
\usepackage{units}
\usepackage{bm}

\def\be{\begin{eqnarray}}
\def\ee{\end{eqnarray}}

\shorttitle{Magnetized inhomogeneous environments near repeating FRB sources}
\shortauthors{Yang et al.} 

\begin{document}

\title{Temporal Scattering, Depolarization, and Persistent Radio Emission from Magnetized Inhomogeneous Environments Near Repeating Fast Radio Burst Sources}

\correspondingauthor{Yuan-Pei Yang} 
\email{ypyang@ynu.edu.cn}

\correspondingauthor{Wenbin Lu} 
\email{wenbinlu@astro.princeton.edu}

\author[0000-0001-6374-8313]{Yuan-Pei Yang}
\affiliation{South-Western Institute for Astronomy Research, Yunnan University, Kunming, Yunnan 650500, China}

\author[0000-0002-1568-7461]{Wenbin Lu}
\affiliation{Department of Astrophysical Sciences, Princeton University, Princeton, NJ 08544, USA}

\author[0000-0002-0475-7479]{Yi Feng}
\affiliation{Zhejiang Lab, Hangzhou, Zhejiang 311121, China}

\author[0000-0002-9725-2524]{Bing Zhang}
\affiliation{Nevada Center for Astrophysics, University of Nevada, Las Vegas, NV 89154, USA}
\affiliation{Department of Physics and Astronomy, University of Nevada, Las Vegas, NV 89154, USA}

\author[0000-0003-3010-7661]{Di Li}
\affiliation{National Astronomical Observatories, Chinese Academy of Sciences, Beijing 100101, China}
\affiliation{University of Chinese Academy of Sciences, Beijing 100049, China}

\begin{abstract} 
Some repeating fast radio burst (FRB) sources exhibit complex polarization behaviors, including frequency-dependent depolarization, variation of rotation measure (RM), and oscillating spectral structures of polarized components. 
Very recently, \citet{Feng22} reported that active repeaters exhibit conspicuous frequency-dependent depolarization and a strong correlation between RM scatter ($\sigma_{\rm RM}$) and the temporal scattering time ($\tau_{\rm s}$), $\sigma_{\rm RM}\propto\tau_{\rm s}^{1.0\pm0.2}$, both of which can be well described by multi-path propagation through a magnetized inhomogeneous plasma screen. 
This observation strongly suggests that the temporal scattering and RM scatter originate from the same region. Besides, a particular finding of note in \citet{Feng22} is that the FRBs with compact persistent radio sources (PRS) tend to have extreme $\sigma_{\rm RM}$. 
In this work, we focus on some theoretical predictions on the relations among temporal scattering, depolarization by RM scatter, and PRS contributed by the magnetized plasma environment close to a repeating FRB source.
The behaviors of the RM scatter imply that the magnetized plasma environment is consistent with a supernova remnant or a pulsar wind nebula, and the predicted $\sigma_{\rm RM}$-$\tau_{\rm s}$ relation is $\sigma_{\rm RM}\propto\tau_{\rm s}^{(0.54-0.83)}$ for different astrophysical scenarios. 
We further make a general discussion on PRS that does not depend on specific astrophysical scenarios.
We show that the specific luminosity of a PRS should have a positive correlation with the RM contributed by the plasma screen. This is consistent with the observations of FRB 121102 and FRB 190520B.

\end{abstract} 


\section{Introduction} 

Fast radio bursts (FRBs) are cosmological radio transients with millisecond durations.  Since the first FRB (FRB 010724, ``Lorimer burst'') was discovered in 2007 \citep{Lorimer07}, hundreds of FRB sources have been detected, dozens of which are repeaters \citep[e.g.,][]{CHIME21}. Recently, a Galactic FRB 200428 was detected to be associated with SGR J1935+2154 \citep{Bochenek20,CHIME20,Li20,Mereghetti20,Ridnaia20,Tavani20}, which suggests that at least some FRBs originate from magnetars born from core collapse of massive stars \citep[e.g.,][]{Popov13,Katz16,Murase16,Beloborodov17,Kumar17,Yang18,Yang21,Metzger19,Lu20,Margalit20,Wadiasingh20,Zhangb21,Wangwy21}. However, FRB 20200120E was found to be in a globular cluster of a nearby galaxy, M81 \citep{Bhardwaj21,Kirsten21}. This is in tension with scenario that invokes active magnetars with age $\lesssim 10~{\rm kyr}$ formed in core-collapse supernovae \citep{Kremer21,Lu21}, and suggests that FRBs might origin from magnetars formed in compact binary mergers \citep{Margalit19,Wang20e,Zhong20,Zhao21}. Therefore, the physical origin of FRBs is still not well constrained from the data \citep[e.g.,][]{Petroff19,Cordes19,Zhang20,Xiao21}. 
The growing FRB detections start to shed lights onto the diversity among the phenomena. The repeaters presented in the first CHIME FRB catalog  have relatively larger widths and narrower bandwidth compared with one-off FRBs \citep{Pleunis21}. The behaviors of fluence with respect to peak flux exhibit statistically significant differences between bursts with long and short durations \citep{Li21c}.  Multiple origins for the FRB population seem increasingly likely.

The first known repeater, FRB 121102, possesses numerous interesting properties including: 1) a bright persistent radio counterpart with a luminosity of $\nu L_{\nu}\sim10^{39}~{\rm erg~s^{-1}}$ at $\nu\sim10~{\rm GHz}$ that is spatially coincident with the FRB source \citep{Chatterjee17,Chen22}; 2) a large rotation measure (RM), $|{\rm RM}|\sim10^5~{\rm rad~m^{-2}}$ \citep{Michilli18}, with a decreasing trend of evolution over the period of a few years \citep{Hilmarsson21}; 3) a large dispersion measure (DM) contribution ($55~{\rm pc~cm^{-3}}\lesssim{\rm DM_{host}}\lesssim225~{\rm pc~cm^{-3}}$) from its host galaxy \citep{Tendulkar17}; and 4) a high burst rate and a bimodal energy distribution with time evolution \citep{Lid21}. These properties imply that  FRB 121102 have a magneto-ionic environment and an active central engine. 
When relativistic electrons hold a good proportion of the electron population in a dense magnetized plasma, bright persistent radio emission would be generated by synchrotron radiation \citep{Murase16,Metzger17,Kashiyama17,Margalit18,Yang20a}\footnote{See, however, \citet{Yang16} and \citet{Liqc20} for an alternative interpretation that the persistent radio emission might be generated by synchrotron-heating of a synchrotron nebula by radio bursts themselves.}.

In addition to FRB 121102, some repeaters recently studied by the Five-hundred-meter Aperture Spherical radio Telescope (FAST, \citealt{Li19}) also exhibit signs of complex magnetized plasma environments. FRB 190520B  \citep{Niu21}, first discovered through drift scans of the Commensal Radio Astronomy FAST Survey (CRAFTS, \citealt{Li18}), is co-located with a compact, persistent radio source (PRS) with luminosity similar to that of FRB 121102, and its host is a dwarf galaxy at $z=0.241$ with high specific star formation rate. The estimated DM contribution from the host is ${\rm DM_{host}}\simeq900~{\rm pc~cm^{-3}}$, nearly an order of magnitude higher than those of other FRBs, which might be explained by a supernova remnant \citep[e.g.,][]{Zhao21b,Katz22}. 

Another active repeater FRB 20201124A \citep{Lanman21,Kumar21,Nimmo21,Hilmarsson21} is found in a Milky-Way-sized, metal-rich, barred-spiral host galaxy at $z=0.098$ \citep{Xu21,Fong21,Ravi21,Piro21}. This repeater indicates a significant, irregular, variation of the Faraday rotation over 36 days. Some bursts appear to have circular polarization up to 75\% \citep{Hilmarsson21b,Kumar21,Xu21}. In particular, the frequency spectra of both circular polarization and linear polarization of some bursts with moderate circular polarization show clear oscillating structures \citep{Xu21}, which might originate from the polarized absorption or the Faraday conversion mechanism \citep{Melrose04,Li22}.

On the other hand, turbulence generally exists in complex magnetized plasma environments, leading to temporal scattering, scintillation, depolarization, etc. The temporal scattering time of some FRBs is much longer than that of radio pulsars at high Galactic latitudes \citep{Cordes16}. Meanwhile, the lack of any correlation between scattering time and DM of FRBs implies that the intergalactic medium cannot account for both scattering time and DM\footnote{This is different from radio pulsars. For radio pulsars, a relation between scattering time and DM has been established \citep{Cordes16}, which implies that the scattering time and DM are contributed by the same source.}, and a detailed analysis for the turbulence effect of the intergalactic medium could be found in \citet{Beniamini20}.
\citet{Qiu20} studied the profiles of some FRBs detected by Australian Square Kilometre Array Pathfinder (ASKAP). Five FRBs were identified with evidence of millisecond pulse broadening caused by scattering in an inhomogeneous plasma, and they suggested that the temporal scattering could be caused by the interstellar medium or near-source plasma in the host galaxy. 
Theoretically, \citet{Xu16} examined some possible density fluctuation turbulence models, and found that a short-wave-dominated power-law density spectrum can interpret the scattering timescale of FRBs.
\citet{Simard21} constrained the turbulence properties by comparing the measurements of FRB scattering with the optical recombination-line tracers of their host environments. In a cold magnetized plasma or a relativistic plasma, Faraday conversion could occur when an FRB propagates in the medium, leading to conversion of linearly polarized emission  to circularly polarized emission \citep{Vedantham19,Gruzinov19}. In a recent study, \citet{Beniamini21} systematically studied the observed polarization properties of an FRB propagating in a magnetized plasma screen via multi-path propagation. 

More recently, \citet{Feng22} reported new polarization measurements of five active repeaters, including FRB 121102, FRB 190520B, FRB 190303, FRB 190417, and FRB 20201124A with FAST and Green Bank Telescope (GBT). These bursts exhibit conspicuous frequency-dependent linear polarization fraction that can be well described by RM scatter, $\sigma_{\rm RM}$ (see Section \ref{depolarization} for a detailed discussion about the depolarization by RM scatter). 
Furthermore, the scattering time $\tau_{\rm s}$ of these bursts show a strong correlation with RM scatter\footnote{Notice that the relation in Figure 4 of \citet{Feng22} is $\tau_{\rm s}\propto\sigma_{\rm RM}^{0.81\pm0.16}$, and the temporal scattering times of all FRBs have been scaled to $\nu\sim 1.3~{\rm GHz}$. For the same data, one has $\sigma_{\rm RM}\propto\tau_{\rm s}^{1.03\pm0.21}$.}, $\sigma_{\rm RM}\propto\tau_{\rm s}^{1.0\pm0.2}$,
which implies that $\sigma_{\rm RM}$ and $\tau_{\rm s}$ likely originate from the same environment. 

In this work, we propose that temporal scattering, depolarization by RM scatter, and persistent radio emission all originate from the magnetized plasma environment near a FRB source, and predict the relations among them. The paper is organized as follows. We consider that an FRB propagates in a magneto-ionic inhomogeneous plasma screen, and calculate temporal scattering and depolarization by RM scatter in Section \ref{sec2}. The persistent radio emission from the magnetized plasma screen is analyzed in Section \ref{sec3}. The results are discussed and summarized in Section \ref{sec4}. 

\section{Temporal scattering and depolarization by RM scatter from a magnetized plasma screen}\label{sec2}  

\begin{figure}[]
    \centering
	\includegraphics[width = 1.2\linewidth, trim = 120 100 0 100, clip]{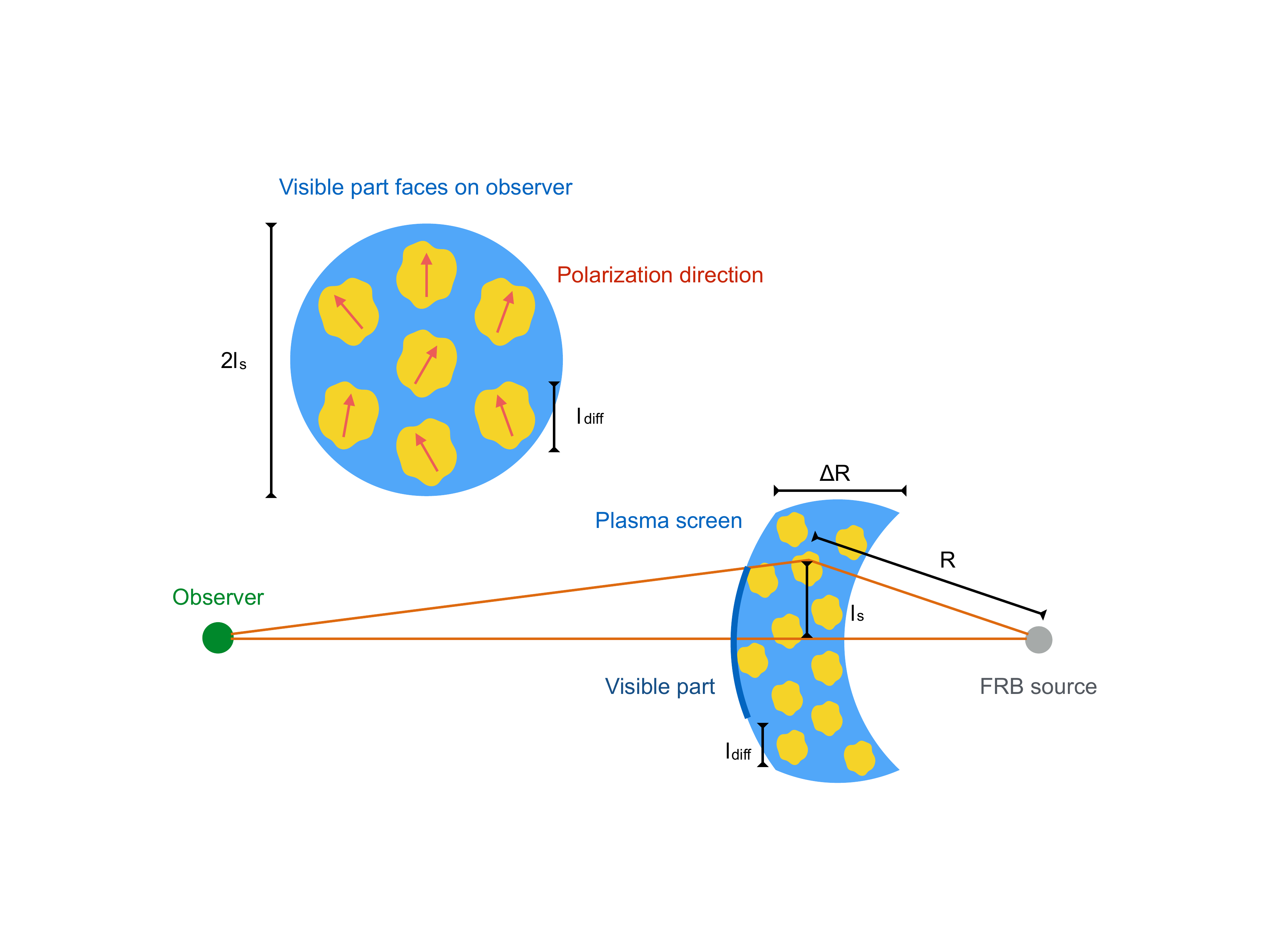}
    \caption{Schematic configuration of an FRB propagating in a magnetized inhomogeneous plasma screen. The yellow regions correspond to clumps with diffractive lengthscale $l_{\rm diff}$. The top left corner shows the polarization directions of electromagnetic waves from each clump, when the visible part of the plasma screen faces the observer.}\label{fig1} 
\end{figure}

\subsection{Temporal scattering}
We consider a power-law spectrum of  electron density fluctuations in a magnetized plasma screen with thickness $\Delta R$ satisfying
\be
P(k)=C_N^2k^{-\beta},~~~{\rm for}~2\pi L^{-1}\lesssim k\lesssim 2\pi l_0^{-1},\label{power}
\ee
where $k=2\pi/l$ is the spatial wavenumber, $L$ and $l_0$ are the outer and inner scales delineating the inertial range of the turbulence, respectively, $\beta$ is the spectral index of the three-dimensional power spectrum, and Kolmogorov turbulence has $\beta=11/3$.
Using the normalization of the power spectrum $\int P(\overrightarrow{k})d\overrightarrow{k}=\delta n_{e,0}^2$, one has \citep[e.g.,][]{Xu17}
\be
C_N^2\simeq
\left\{
\begin{aligned}
&\frac{3-\beta}{2(2\pi)^{4-\beta}}l_0^{3-\beta}\delta n_{e,0}^2,~&{\rm for}~\beta<3,\\
&\frac{\beta-3}{2(2\pi)^{4-\beta}}L^{3-\beta}\delta n_{e,0}^2,~&{\rm for}~\beta>3, 
\end{aligned}
\right.
\ee
where $\delta n_{e,0}^2$ is the total mean-squared density fluctuation.
According to Eq.(\ref{power}), for $l_0<l<L$ the electron density fluctuation $\delta n_{e,l}$ at scale $l$ is
\be
\delta n_{e,l}^2\sim 4\pi C_N^2 k^{3-\beta}\sim
\left\{
\begin{aligned}
&\delta n_{e,0}^2\left(\frac{l}{l_0}\right)^{\beta-3},~&{\rm for}~\beta<3,\\
&\delta n_{e,0}^2\left(\frac{l}{L}\right)^{\beta-3},~&{\rm for}~\beta>3. 
\end{aligned}
\right.\label{nl}
\ee
When the radio wave propagates in a turbulent medium, the fluctuating refractive indices introduce random phase fluctuations to the wavefront, as shown in Figure \ref{fig1}. The ``phase structure function'' is used to represent the mean-squared phase difference between two points separated by $l$, $D_\phi(\overrightarrow{l})\equiv\left<[\phi(\overrightarrow{x}+\overrightarrow{l})-\phi(\overrightarrow{x})]^2\right>$. If the turbulence in the plasma screen is isotropic and $L\ll \Delta R$, the phase structure function is given by \citep{Coles87,Rickett90,Xu17}
\be
D_\phi(l)\simeq
\left\{
\begin{aligned}
&f_{1,\alpha}\pi^2 r_e^2\lambda^2C_N^2\Delta Rl_0^{\beta-2}\left(\frac{l}{l_0}\right)^2,~&{\rm for}~l\lesssim l_0,\\
&f_{2,\alpha}\pi^2 r_e^2\lambda^2 C_N^2\Delta Rl_0^{\beta-2}\left(\frac{l}{l_0}\right)^{\beta-2},~&{\rm for}~l\gtrsim l_0, 
\end{aligned}
\right.
\ee
where $\lambda$ is the wavelength of the electromagnetic wave, $f_{1,\alpha}=\Gamma(1-\alpha/2)$, $f_{2,\alpha}=[\Gamma(1-\alpha/2)/\Gamma(1+\alpha/2)](8/\alpha 2^{\alpha})$, and $\alpha=\beta-2$. 
For the Kolmogorov turbulence with $\alpha=5/3~(\beta=11/3)$, one has $f_{1,\alpha}=5.6$ and $f_{2,\alpha}=8.9$, respectively. 
We define $l_{\rm diff}$ as the diffractive lengthscale that represents the transverse separation for which the root-mean-squared difference is equal to 1 rad, leading to $D_\phi(l_{\rm diff})=1$, one has
\be
l_{\rm diff}=
\left\{
\begin{aligned}
&(f_{1,\alpha}\pi^2 r_e^2\lambda^2l_0^{\beta-4}C_N^2\Delta R)^{-\frac{1}{2}},~&{\rm for}~l_{\rm diff}<l_0,\\
&(f_{2,\alpha}\pi^2 r_e^2\lambda^2 C_N^2\Delta R)^{\frac{1}{2-\beta}},~&{\rm for}~l_{\rm diff}>l_0, 
\end{aligned}
\right.\label{ldiff}
\ee

For the multi-path propagation as shown in Figure \ref{fig1}, the scattering angle of the electromagnetic waves is approximately $\theta_{\rm s}\simeq\lambda/2\pi l_{\rm diff}$.
Therefore, the transverse scale of the visible part is
\be
l_{\rm s}(\lambda)=\theta_{\rm s} R\simeq\frac{\lambda R}{2\pi l_{\rm diff}},\label{lsca}
\ee
where $R$ is the distance from the plasma screen to the source.
The path-length difference between two rays is $\Delta s\simeq R(1-\cos\theta_{\rm s})\simeq R\theta_{\rm s}^2/2$ for $\theta_{\rm s}\ll1$. The temporal scattering time could be then estimated by\footnote{Notice that due to the expansion of the universe, the observed wavelength is $\lambda_{\rm obs}=(1+z)\lambda$, where $\lambda$ is the wavelength at the plasma screen. Considering that most repeaters are at low redshifts, we ignore the redshift correction in the following discussion.} 
\be
\tau_{\rm s}(\lambda)\simeq\frac{l_{\rm s}^2}{2Rc}=\frac{\lambda^2R}{8\pi^2 cl_{\rm diff}^2}.\label{tau}
\ee
Using Eq.(\ref{ldiff}) and $\Delta R\sim R$, the temporal scattering time satisfies
\be
\tau_{\rm s}(\lambda)\propto
\left\{
\begin{aligned}
&\delta n_{e,0}^2R^2\lambda^4,&l_{\rm diff}<l_0,\\
&\delta n_{e,0}^{\frac{4}{\beta-2}}R^{\frac{\beta}{\beta-2}}\lambda^{\frac{2\beta}{\beta-2}},&l_{\rm diff}>l_0. 
\end{aligned}
\right.\label{tau1}
\ee

For a certain plasma screen, scintillation and temporal scattering occur together, and have a relation of $\Delta \nu_{\rm sci}=1/(2\pi\tau_{\rm s})$, where $\Delta \nu_{\rm sci}$ is the scintillation bandwidth.
In general, scintillation refers to spectral modulations, and temporal scattering refers to temporal broadening of pulses due to the multi-path propagation effect. The relation between scintillation and temporal scattering is due to the following reason: if the optical path difference is $\sim c\tau_{\rm s}$, the phase difference is $\sim(2\pi/\lambda)c\tau_{\rm s}=2\pi\nu \tau_{\rm s}$; thus the phase of the interference waves change by $\sim 1~{\rm rad}$ in the bandwidth of $\Delta\nu_{\rm sci}=1/({2\pi}\tau_{\rm s}$).  
The value of $\Delta \nu_{\rm sci}\simeq160~{\rm Hz}(\tau_{\rm s}/1~{\rm ms})^{-1}$ is much smaller than the observed scintillation bandwidth of $\sim1{\rm MHz}$. 
This result suggests that for extragalactic FRBs, the observed scintillation is mainly contributed by the interstellar medium within the Milky Way, whereas the observed scattering time is more likely contributed by circumburst medium or the interstellar medium in the FRB host galaxy. 

\subsection{Depolarization due to RM scatter}\label{depolarization}
Next, we discuss the depolarization effect of the RM scatter from the magnetized plasma screen. This has been discussed by \citet{Beniamini21}, but here we show that there is a correlation between two measurable quantities, i.e. the scattering time and the depolarization wavelength --- the critical wavelength below which radio waves are depolarized.
We consider that there is a fluctuation in $n_e$ and $B_{\parallel}$ across the lengthscale $l$. In a magneto-ionic cold plasma, the dispersion relation of left and right circularly polarized waves is approximately
\be
\frac{k^2c^2}{\omega^2}\simeq1-\frac{\omega_p^2}{\omega^2} \pm \frac{\omega_p^2\omega_B}{\omega^3},
\ee
for $\omega\gg\omega_B$,
where $\omega_p=(4\pi e^2n_e/m_e)^{1/2}$ is the plasma frequency, $\omega_B=eB_\parallel/m_ec$ is the electron cyclotron frequency. Faraday rotation is related to the parallel component $B_\parallel = B\cos\theta$, and $\theta$ is the angle between the wave-vector and the local magnetic field.
For a certain polarized wave (left or right circular polarized waves) propagating in a homogeneous medium with scale $l$, the phase difference before and after crossing the medium is $\phi_\pm=k_\pm l$ (apart from the trivial light travel time difference). A given path contains on average $N=\Delta R/l$ independent segments, and the Poisson root mean squared (RMS) fluctuations is $\Delta N = (\Delta R/l)^{1/2}$ in the limit of $\Delta N\gg 1$.
Therefore, the total phase perturbation contributed by the plasma screen with an inhomogeneous medium is 
\be
\Delta\phi_\pm(l)&=&\left(\frac{\Delta R}{l}\right)^{1/2}\delta\phi_\pm\nonumber\\
&\simeq&\left(\frac{\Delta R}{l}\right)^{1/2}\frac{\omega l}{c}\left[\frac{\delta(\omega_p^2)}{2\omega^2}\mp\frac{\delta(\omega_p^2\omega_B)}{2\omega^3}\right],
\ee
for $\omega\gg\omega_B,\omega_p$, where $\delta\phi_\pm$ is the phase perturbation of the left/right circular polarized waves in a clump with lengthscale $l$. The perturbation of the Faraday rotation angle after a radio burst propagating across the plasma screen is $\Delta\psi=|\Delta\phi_+-\Delta\phi_-|/2$, i.e.,
\be
\Delta\psi(l)\simeq\left(\frac{\Delta R}{l}\right)^{1/2}\delta {\rm RM}(l)\lambda^2=\frac{2\pi e^3(l\Delta R)^{1/2}}{m_e^2c^2\omega^2}\delta(n_eB_\parallel)_l,\nonumber\\
\ee
where the RM contribution by a clump with scale $l$ is given by
\be
\delta {\rm RM}(l)=\frac{e^3}{2\pi m_e^2c^4} \delta(n_eB_\parallel)_l l,
\ee
and $\delta(n_e B_\parallel)_l$ is on the scale of $l$.
One must notice that the decorrelation lengthscale of the Faraday rotation angle is different from the diffractive scale $l_{\rm diff}$ that reflects the phase structure function, as shown in Figure \ref{fig1}.

Since the rotation angle $\Delta\psi$ increases with $l$, the most important contribution comes from the largest transverse separation $l_{\rm s}$ given by Eq.(\ref{lsca}). 
Let us consider the source to be 100\% linearly polarized. If $\Delta \psi(l_{\rm s}) \gg 1\rm\, rad$, the observed waves at any given time will be the superposition of a large number of patches with random polarization directions. The size of each patch, $l_{\rm PA}$, can be estimated by\footnote{We consider the realistic limit $\omega_B\ll \omega$ (or $B\ll 1\rm\, kG$ for GHz wave) and hence $\Delta \psi(l)\ll \Delta \phi(l)$. Thus, each patch of the size of $l_{\rm PA}$ contains many sub-patches (each of the size of $l_{\rm diff}$) of different phases --- even though the waves contributed by a patch of $l_{\rm PA}$ have the same polarization direction, the phase coherence is lost.} $\Delta \psi(l_{\rm PA})\sim 1\rm\, rad$. 
Then, when we add up the contributions from many $l_{\rm PA}$ patches, the final waves are unpolarized.
Therefore, depolarization occurs at the depolarization wavelength $\lambda_{\rm dep}$ when $\Delta\psi(l_{\rm s})\sim 1~{\rm rad}$. The RM scatter contributed by the plasma screen is estimated by
\be
\sigma_{\rm RM}&\simeq&\left(\frac{\Delta R}{l_{\rm s}}\right)^{1/2}\delta {\rm RM}(l_{\rm s})\nonumber\\
&=&\frac{e^3}{2\pi m_e^2c^4} (l_{\rm s}\Delta R )^{1/2} \delta(n_eB_\parallel)_{l_{\rm s}}\nonumber\\
&=&0.81~{\rm rad~m^{-2}}\left(\frac{\sqrt{l_{\rm s}\Delta R}}{{1~\rm pc}}\right)\left(\frac{\delta(n_eB_\parallel)_{l_{\rm s}}}{1~{\rm cm^{-3}\mu G}}\right),\label{sigma}
\ee
and the depolarization wavelength is
\be
\lambda_{\rm dep}\sim\sigma_{\rm RM}^{-1/2}.\label{lambda}
\ee
We notice that the observed RM scatter $\sigma_{\rm RM}$ should be always less than the absolute value of RM contributed by the magnetized plasma screen. The observed result of $\sigma_{\rm RM}\ll |{\rm RM}|$ \citep{Feng22} implies that a large-scale magnetic field may exist in the screen, or the observed RM is contributed by other regions.
According to Eq.(\ref{tau}) and Eq.(\ref{sigma}), eliminating $l_{\rm s}$ and taking $\Delta R\sim R$, one finally obtains
\be
\sigma_{\rm RM}&\simeq&\frac{e^3}{2\pi m_e^2c^4} \delta(n_eB_\parallel)_{l_{\rm s}}(2R^3 c \tau_{\rm s}(\lambda_{\rm dep}))^{1/4}=1.7~{\rm rad~m^{-2}}\nonumber\\
&\times&\left(\frac{R}{{1~\rm pc}}\right)^{3/4}\left(\frac{\delta(n_eB_\parallel)_{l_{\rm s}}}{{10^3~\rm cm^{-3}\mu G}}\right)\left(\frac{\tau_{\rm s}(\lambda_{\rm dep})}{1~{\rm ms}}\right)^{1/4},
\nonumber\\\label{sigma1}
\ee
where $\tau_{\rm s}(\lambda_{\rm dep})$ is the scattering time at the depolarization wavelength.
Some repeating FRBs studied by \citet{Feng22} have RM scatter values $\sigma_{\rm RM}\gtrsim 1~{\rm rad~m^{-2}}$. This is consistent with the picture that the RM scatter and temporal scattering originate from radio bursts propagating in a inhomogeneous magneto-ionic environment near the source. The above typical parameters of the plasma screen are consistent with the scenario of a supernova remnant or a pulsar wind nebula \citep{Reynolds12,Feng22}. 

\subsection{$\sigma_{\rm RM}-\tau_s$ relations}
In order to obtain the $\sigma_{\rm RM}-\tau_{\rm s}$ relation, we would like to relate $\sigma_{\rm RM}$ to the temporal scattering time $\tau_{\rm s,0}$ at a fixed wavelength $\lambda_0$ for all repeaters, as measured by \citet{Feng22}. According to Eq.(\ref{tau1}) and Eq.(\ref{lambda}), the temporal scattering time at wavelength $\lambda_{\rm dep}$ is given by
\be
\tau_{\rm s}\propto
\left\{
\begin{aligned}
&\sigma_{\rm RM}^{-2}\tau_{\rm s,0},&l_{\rm diff}<l_0,\\
&\sigma_{\rm RM}^{\frac{\beta}{2-\beta}}\tau_{\rm s,0},&l_{\rm diff}>l_0, 
\end{aligned}
\right.\label{tau0}
\ee
Since the lengthscale of the visible part, $l_{\rm s}$, could be larger than the maximum length scale $L$ of turbulence, we will discuss the cases of $l_{\rm s}\lesssim L$ and $l_{\rm s}\gtrsim L$.

(1) $l_0\lesssim l_{\rm s}\lesssim L$: In this case, one has $\delta n_e\ll n_e$ and $\delta B_\parallel \ll B_\parallel$, so that
\be
\delta(n_e B_\parallel)\sim B_{\parallel} \delta n_e.
\ee
Using Eq.(\ref{nl}) and Eq.(\ref{tau}), one further obtains $\delta (n_e B_\parallel)_{l_{\rm s}}\propto B_\parallel \delta n_{e,0} l_{\rm s}^{(\beta-3)/2}\propto B_\parallel  \delta n_{e,0}\tau_{\rm s}^{(\beta-3)/4}R^{(\beta-3)/4}$. 
Using Eq.(\ref{tau0}), one gets
\be
\delta (n_e B_\parallel)_{l_{\rm s}}\propto
\left\{
\begin{aligned}
&B_\parallel  \delta n_{e,0}\sigma_{\rm RM}^{\frac{3-\beta}{2}}\tau_{\rm s,0}^{\frac{\beta-3}{4}}R^{\frac{\beta-3}{4}},&l_{\rm diff}<l_0,\\
&B_\parallel  \delta n_{e,0}\sigma_{\rm RM}^{\frac{\beta(\beta-3)}{4(2-\beta)}}\tau_{\rm s,0}^{\frac{\beta-3}{4}}R^{\frac{\beta-3}{4}},&l_{\rm diff}>l_0. 
\end{aligned}
\right.\nonumber\\\label{nB}
\ee
According to Eq.(\ref{tau1}) Eq.(\ref{sigma1}), Eq.(\ref{tau0}) and Eq.(\ref{nB}), one finally obtains the $\sigma_{\rm RM}$-$\tau_{\rm s,0}$ relation,
\be
\sigma_{\rm RM}\propto
\left\{
\begin{aligned}
&\tau_{\rm s,0}^{\frac{1}{2}}R^{\frac{\beta-4}{2\beta}}B_\parallel^{\frac{2}{\beta}},&l_{\rm diff}<l_0,\\
&\tau_{\rm s,0}^{\frac{2(\beta-2)}{\beta+4}}R^{0}B_\parallel^{\frac{4}{\beta+4}},&l_{\rm diff}>l_0, 
\end{aligned}
\right.\label{sigmatau1a}
\ee 

(2) $l_{\rm s}\gtrsim L$: In this case, one has $\delta n_e\sim \delta n_{e,0}\sim n_e$ and $\delta B_\parallel\sim\delta B_{\parallel,0} \propto B_\parallel$, where $\delta B_{\parallel,0}$ is the total root-mean-squared parallel magnetic field. For the global turbulent magnetic field, $\delta B_{\parallel,0}\sim B_\parallel$. For the magnetic field with turbulent component and large-scale component, $\delta B_{\parallel,0}<B_\parallel$.  Thus, one may have
\be
\delta(n_e B_\parallel)_{l_{\rm s}}\propto B_{\parallel} \delta n_{e,0}.
\label{nB1}
\ee
Using Eq.(\ref{tau1}) Eq.(\ref{sigma1}), Eq.(\ref{tau0}) and Eq.(\ref{nB1}), the $\sigma_{\rm RM}$-$\tau_{\rm s,0}$ relation becomes
\be
\sigma_{\rm RM}\propto
\left\{
\begin{aligned}
&\tau_{\rm s,0}^{\frac{1}{2}}R^{-\frac{1}{6}}B_\parallel^{\frac{2}{3}},&l_{\rm diff}<l_0,\\
&\tau_{\rm s,0}^{\frac{(\beta-2)(\beta-1)}{5\beta-8}}R^{\frac{\beta^2-5\beta+6}{8-5\beta}}B_\parallel^{\frac{4(\beta-2)}{5\beta-8}},&l_{\rm diff}>l_0.
\end{aligned}
\right.\label{sigmatau1b}
\ee
The $\sigma_{\rm RM}$-$\tau_{\rm s}$ relation given by Eq.(\ref{sigmatau1a}) and Eq.(\ref{sigmatau1b}) involves several parameters connected to the nature of the screens, including screen radius $R$, line-of-sight component of magnetic field $B_\parallel$, turbulence inner scale $l_0$, etc, which may vary between different FRBs, and the variations of these parameters would affect the scatter of the $\sigma_{\rm RM}$-$\tau_{\rm s}$ relation. The larger the variation of these parameters, the larger the scatter of the $\sigma_{\rm RM}$-$\tau_{\rm s}$ relation.
Further more, the dependence on  distance $R$ is negligible and the only unknown is $B_\parallel$. Because the sources with stronger turbulent fluctuations are expected to have a higher magnetic field strength, we expect the scaling to be steeper than $\sigma_{\rm RM} \propto \tau_{\rm s,0}^{1/2}$. 

In the following, we propose to estimate the parallel component of the magnetic field by the fluctuations of RM and DM \citep{Katz18,Katz21},
\be
B_\parallel&\simeq&\frac{2\pi m_e^2c^4}{e^3}\frac{\Delta{\rm RM}}{\Delta{\rm DM}} \nonumber\\
&=& 1.2~\mathrm{mG}\, \left(\frac{\Delta\rm RM}{\rm 10^3\, rad\,m^{-2}}\right) \left(\frac{\Delta \rm DM}{1\rm pc\, cm^{-3}}\right)^{-1}.
\ee 
Because the large-scale plasma, e.g., the interstellar or intergalactic medium, contributes to nearly time-invariant RM and DM, the fluctuations $\Delta {\rm RM}$ and $\Delta {\rm DM}$ on timescales less than a few years are expected to originate from either the time evolution of the local plasma or the proper motion of the source with respect to the local plasma \citep{Yang17, Piro18}.

If one conducts an extensive monitoring campaign on a repeater source, it is possible to measure the root-mean-squared variation of $\left<\Delta{\rm RM}\right>_t$ and $\left<\Delta{\rm DM}\right>_t$, where $\left<...\right>_t$ denotes an ensemble average of multiple measurements of different $\Delta {\rm RM}$ and $\Delta {\rm DM}$ at a time separation of $t$. For example, FRB 121102 had an averaged increase of DM by $\sim1\rm\, pc\,cm^{-3}$ per year \citep{Hessels19} and an average decrease of RM by $\sim 10^4\rm\, rad\, m^{-2}$ per year \citep{Hilmarsson21}, so we infer $B_{\parallel}\sim 10\rm\, mG$. Another source FRB 20201124A had strong RM fluctuations of $\Delta\rm RM\sim 200\rm\, rad\,m^{-2}$ on a timescale of 10 days \citep{Xu21} whereas its DM fluctuation is not well measured (due to the long scattering time) but constrained to be $\Delta\rm DM\lesssim 3\rm\, pc\,cm^{-3}$ on a similar timescale \citep{Xu21}. For this source, we can infer $B_{\parallel}\gtrsim 0.1\rm\, mG$.
Even if good measurements of $\Delta {\rm RM}$ and $\Delta {\rm DM}$ are not available, for a source with a large ${\rm RM}\gtrsim 500~{\rm rad~m^{-2}}$, the main contributor to RM would possibly be from the local plasma. Meanwhile, the upper limit on the local DM may be inferred from the host galaxy redshift and Galactic DM contributions, so it is still possible to estimate a lower limit of $B_\parallel$ in the local plasma. Therefore, the relations of Eq.(\ref{sigmatau1a}) and Eq.(\ref{sigmatau1b}) can be tested by observations. 

Since the magnetic fields near most repeaters have not been measured by the above method, in the following we consider the magnetic field---density relation satisfies
\be
B=An_e^\kappa
\ee
Different sources have the same $\kappa$ but different values of $A$. One may consider two cases: (1) if $l_0\lesssim l_{\rm s}\lesssim L$, according to Eq.(\ref{tau1}) and Eq.(\ref{sigmatau1a}), one obtains
\be
\sigma_{\rm RM}\propto
\left\{
\begin{aligned}
&\tau_{\rm s,0}^{\frac{\beta+2\kappa}{2\beta}}R^{\frac{\beta-4\kappa-4}{2\beta}},&l_{\rm diff}<l_0,\\
&\tau_{\rm s,0}^{\frac{(\beta-2)(\kappa+2)}{\beta+4}}R^{\frac{-\beta\kappa}{\beta+4}},&l_{\rm diff}>l_0.
\end{aligned}
\right.\label{case1}
\ee
We define the variation range of $A$ for different sources as $\delta A$, using Eq.(\ref{sigmatau1a}), the scatter contributed by $\delta A$ of the above relation is
\be
\frac{\delta\sigma_{\rm RM}}{\bar\sigma_{\rm RM}}=
\left\{
\begin{aligned}
&\left(\frac{\delta A}{\bar A}\right)^{\frac{2}{\beta}},&l_{\rm diff}<l_0,\\
&\left(\frac{\delta A}{\bar A}\right)^{\frac{4}{\beta+4}},&l_{\rm diff}>l_0.
\end{aligned}
\right.\label{un1}
\ee
(2) if $l_{\rm s}\gtrsim L$, according to Eq.(\ref{tau1}) and Eq.(\ref{sigmatau1b}), one obtains
\be
\sigma_{\rm RM}\propto
\left\{
\begin{aligned}
&\tau_{\rm s,0}^{\frac{2\kappa+3}{6}}R^{\frac{-(4\kappa+1)}{6}},&l_{\rm diff}<l_0,\\
&\tau_{\rm s,0}^{\frac{(\beta-2)(\kappa\beta+\beta-2\kappa-1)}{5\beta-8}}R^{\frac{(\kappa+1)\beta^2-(2\kappa+5)\beta+6}{8-5\beta}},&l_{\rm diff}>l_0. 
\end{aligned}
\right.\nonumber\\\label{case2}
\ee
Using Eq.(\ref{sigmatau1b}),the scatter of the above relation is
\be
\frac{\delta\sigma_{\rm RM}}{\bar\sigma_{\rm RM}}=
\left\{
\begin{aligned}
&\left(\frac{\delta A}{\bar A}\right)^{\frac{2}{3}},&l_{\rm diff}<l_0,\\
&\left(\frac{\delta A}{\bar A}\right)^{\frac{4(\beta-2)}{5\beta-8}},&l_{\rm diff}>l_0.
\end{aligned}
\right.\label{un2}
\ee
In the following discussion, we will discuss three different astrophysical scenarios.

\subsubsection{Magnetized plasma with energy equipartition}\label{sec21}
If the magnetized plasma screen roughly satisfies the energy equipartition between magnetic energy and kinetic energy of thermal electrons, one has
\be
n_ek_BT \sim \frac{B^2}{8\pi},\label{balance}
\ee
where $k_B$ is the Boltzmann constant and $T$ is the plasma temperature.
(1) if $l_0\lesssim l_{\rm s}\lesssim L$, according to Eq.(\ref{case1}), one obtains
\be
\sigma_{\rm RM}\propto
\left\{
\begin{aligned}
&\tau_{\rm s,0}^{\frac{\beta+1}{2\beta}}R^{\frac{\beta-6}{2\beta}},&l_{\rm diff}<l_0,\\
&\tau_{\rm s,0}^{\frac{5(\beta-2)}{2(\beta+4)}}R^{\frac{-\beta}{2(\beta+4)}},&l_{\rm diff}>l_0. 
\end{aligned}
\right.\label{sigmatau2a}
\ee
(2) if $l_{\rm s}\gtrsim L$, according to Eq.(\ref{case2}), one obtains
\be
\sigma_{\rm RM}\propto
\left\{
\begin{aligned}
&\tau_{\rm s,0}^{\frac{2}{3}}R^{-\frac{1}{2}},&l_{\rm diff}<l_0,\\
&\tau_{\rm s,0}^{\frac{(\beta-2)(3\beta-4)}{2(5\beta-8)}}R^{\frac{3(\beta-2)^2}{2(8-5\beta)}},&l_{\rm diff}>l_0. 
\end{aligned}
\right.\label{sigmatau2b}
\ee
This predicts $\sigma_{\rm RM}\propto\tau_{\rm s,0}^{(0.54-0.67)}$ for Kolmogorov turbulence with $\beta=11/3$, which is shallower than the observed relation given by \citet{Feng22}. 
In particular, if the magnetized plasma screen is photo-ionized with typical temperature of $T\sim 10^4\rm\, K$, the scatter of the above relations would be very small. 

\subsubsection{Magnetic freezing plasma}\label{sec23}
We next consider that turbulent plasma satisfies the condition of the magnetic frozen, which might be satisfied when the magnetic reconnection is not significant. Then one obtains
\be
B=An_e^{2/3}
\ee
(1) if $l_0\lesssim l_{\rm s}\lesssim L$, according to Eq.(\ref{case1}), one obtains
\be
\sigma_{\rm RM}\propto
\left\{
\begin{aligned}
&\tau_{\rm s,0}^{\frac{3\beta+4}{6\beta}}R^{\frac{3\beta-20}{6\beta}},&l_{\rm diff}<l_0,\\
&\tau_{\rm s,0}^{\frac{8(\beta-2)}{3(\beta+4)}}R^{\frac{-2\beta}{3(\beta+4)}},&l_{\rm diff}>l_0. 
\end{aligned}
\right.\label{sigmatau4a}
\ee
(2) if $l_{\rm s}\gtrsim L$, according to Eq.(\ref{case2}), one obtains
\be
\sigma_{\rm RM}\propto
\left\{
\begin{aligned}
&\tau_{\rm s,0}^{\frac{13}{18}}R^{-\frac{11}{18}},&l_{\rm diff}<l_0,\\
&\tau_{\rm s,0}^{\frac{(\beta-2)(5\beta-7)}{3(5\beta-8)}}R^{\frac{5\beta^2-19\beta+18}{3(8-5\beta)}},&l_{\rm diff}>l_0. 
\end{aligned}
\right.\label{sigmatau4b}
\ee
This predicts $\sigma_{\rm RM}\propto\tau_{\rm s,0}^{(0.58-0.72)}$ for Kolmogorov turbulence with $\beta=11/3$. The scatters of the above relations depends on the range of $A$ in the special astrophysical environments, see Eq.(\ref{un1}) and Eq.(\ref{un2}).

\subsubsection{Shock compressed magnetized plasma}\label{sec22}

We further consider the case that the magnetized turbulent plasma is from a shocked medium, and assume that the interstellar media as the upstream of the shocks. 
The Rankine---Hugoniot relation for non-relativistic perpendicular shock requires\footnote{Notice that the shock surface may not necessarily be perpendicular to the line of sight. This may lead to a significant contribution of $B_\parallel$.}
\be
\frac{B}{n_e}\sim \left(\frac{B}{n_e}\right)_{\rm ISM}. \label{shock}
\ee
Again according to Eq.(\ref{case1}) and Eq.(\ref{case2}), we consider two cases: (1) if $l_0\lesssim l_{\rm s}\lesssim L$, one obtains
\be
\sigma_{\rm RM}\propto
\left\{
\begin{aligned}
&\tau_{\rm s,0}^{\frac{\beta+2}{2\beta}}R^{\frac{\beta-8}{2\beta}},&l_{\rm diff}<l_0,\\
&\tau_{\rm s,0}^{\frac{3(\beta-2)}{\beta+4}}R^{\frac{-\beta}{\beta+4}},&l_{\rm diff}>l_0. 
\end{aligned}
\right.\label{sigmatau3a}
\ee
(2) if $l_{\rm s}\gtrsim L$, one obtains
\be
\sigma_{\rm RM}\propto
\left\{
\begin{aligned}
&\tau_{\rm s,0}^{\frac{5}{6}}R^{-\frac{5}{6}},&l_{\rm diff}<l_0,\\
&\tau_{\rm s,0}^{\frac{(2\beta-3)(\beta-2)}{5\beta-8}}R^{\frac{2\beta^2-7\beta+6}{8-5\beta}},&l_{\rm diff}>l_0. 
\end{aligned}
\right.\label{sigmatau3b}
\ee
This predicts $\sigma_{\rm RM}\propto\tau_{\rm s,0}^{(0.65-0.83)}$ for Kolmogorov turbulence with $\beta=11/3$. This theoretically predicted relation is closer to the observed relation  \citep{Feng22} than the prediction of the model invoking magnetized turbulent plasma with energy equipartition or with magnetic frozen condition, as discussed in Section \ref{sec21} and Section \ref{sec23}, respectively. At last, we discuss the scatters of the above relations. In general, the interstellar medium satisfies the energy equipartition condition, thus one has $A=B_{\rm ISM}/n_{\rm ISM}\propto n_{\rm ISM}^{-1/2}$. We assume that different FRB sources have $n_{\rm ISM}$ varying with three orders of magnitude, then $\delta A/\bar A$ varies with one to two orders of magnitude.
According to Eq.(\ref{un1}) and Eq.(\ref{un2}), the relative uncertainty $\delta\sigma_{\rm RM}/\bar \sigma_{\rm RM}$ is with one order of magnitude, which is approximately consistent with the scatter of the observed relation (see Figure 4 of \citet{Feng22}).

\section{Persistent radio emission from a magnetized plasma screen}\label{sec3} 

The FRB sources with large RM and $\sigma_{\rm RM}$ values imply a dense and magnetized environment, which could produce synchrotron radiation if relativistic electrons make up a significant fraction of the plasma energy density, powering a bright PRS \citep[e.g.,][]{Murase16,Metzger17,Margalit18,Yang20a}.
We consider that the electron distribution has a thermal component satisfying three-dimensional Maxwell distribution in the low-energy regime and a non-thermal component in the high-energy regime\footnote{The distribution given by Eq.(\ref{electrondistribution}) is consistent with the particle-in-cell simulations of particle acceleration in relativistic shock \citep{Spitkovsky08}. However, one should notice that the Maxwell distribution in \citet{Spitkovsky08} is two-dimensional, leading to $n_e(\gamma)\propto\gamma$ for $\gamma\lesssim\gamma_{\rm th}$. For the three-dimensional Maxwell distribution, one has $n_e(\gamma)\propto\gamma^2$ for $\gamma\lesssim\gamma_{\rm th}$ \citep{Rybicki86,Kato07}.}, i.e.
\be
n_{e}(u)\equiv\frac{dn_e}{du}\simeq
\left\{
\begin{aligned}
&\frac{n_e}{u_{\rm th}}\left(\frac{u}{u_{\rm th}}\right)^{2},~&u<u_{\rm th},\\
&\frac{n_e}{u_{\rm th}}\left(\frac{u}{u_{\rm th}}\right)^{-p},~&u>u_{\rm th}, 
\end{aligned}
\right.\label{electrondistribution}
\ee
where $u=\sqrt{\gamma^2-1}$ is the dimensionless four velocity, $u_{\rm th}=\sqrt{\gamma_{\rm th}^2-1}$ is the thermal four velocity with $\gamma_{\rm th}=kT/m_ec^2+1$ as the thermal Lorentz factor, $p$ is the distribution index of the power-law component. The RM contribution from relativistic electrons is suppressed by a factor of $\gamma^2$ due to the relativistic mass $m_e\rightarrow\gamma m_e$. 
Therefore, the RM contributed by the plasma screen with the above electron distribution is approximately given by \citep[e.g., Appendix of][]{Quataert00}
\be
{\rm RM}
\simeq\frac{e^3}{2\pi m_e^2c^4}\frac{n_e B_\parallel}{\gamma_{\rm th}^2} \Delta R.\label{RM}
\ee
If the thermal component is non-relativistic, i.e., $\gamma_{\rm th}\sim1$, the RM would be mainly contributed by the non-relativistic electrons, leading to the classical resluts.

For a single electron, the synchrotron radiation power is $P=(4/3)\sigma_{\rm T}c\gamma^2 B^2/8\pi$, and the characteristic synchrotron frequency is $\nu=\gamma^2eB/2\pi m_ec$.
Thus, the spectral radiation power satisfies $P_\nu\simeq P/\nu=m_ec^2\sigma_{\rm T}B/3e$, which depends on $B$ only. We define $\zeta_e$ as the fraction of electrons that radiate synchrotron emission in the GHz band. The electrons emitting synchrotron radiation in the GHz band is required to have a Lorentz factor 
\be
\gamma_{\rm GHz}\sim\left(\frac{2\pi m_ec\nu}{eB}\right)^{1/2}\simeq600\left(\frac{\nu}{1~{\rm GHz}}\right)^{1/2}\left(\frac{B}{1~{\rm mG}}\right)^{-1/2},\label{gammaghz}
\ee
then the fraction $\zeta_e$ is approximately given by
\be
\zeta_e\sim\frac{\gamma_{\rm GHz}n_e(\gamma_{\rm GHz})}{n_e}\sim\left(\frac{\gamma_{\rm GHz}}{\gamma_{\rm th}}\right)^{1-p}
\ee
for $\gamma_{\rm GHz}>\gamma_{\rm th}$. 
The total number of relativistic electrons is approximately $N_e\sim4\pi R^2\Delta R\zeta_e n_e/3$ for $\Delta R\sim R$. 
The specific luminosity of synchrotron radiation is 
\be
L_{\nu}&=&N_e P_\nu = \frac{64\pi^3}{27}\zeta_e\gamma_{\rm th}^2m_ec^2R^2\left|{\rm RM}\right|\nonumber \\
& \simeq&5.7\times10^{29}~{\rm erg~s^{-1}~Hz^{-1}} \nonumber \\
&\times&\left(\frac{\zeta_e\gamma_{\rm th}^2}{0.01}\right) \left(\frac{\left|{\rm RM}\right|}{10^3~{\rm rad~m^{-2}}}\right)\left(\frac{R}{1~{\rm pc}}\right)^2. \label{lum}
\ee
Since $p\sim2$ is satisfied in most astrophysical scenarios of particle acceleration, one has
$\zeta_e\gamma_{\rm th}^2\sim\gamma_{\rm th}^3\gamma_{\rm GHz}^{-1}$.
Therefore, according to Eq.(\ref{gammaghz}), a small value of $\zeta_e\gamma_{\rm th}^2$ requires $\gamma_{\rm th}\lesssim{\rm a~few}$, and the thermal component could be non-relativistic.
At last, we are also interested in the specific luminosity related to $\sigma_{\rm RM}$.
We define $\xi_{nB}=\delta(n_eB_\parallel)_{l_{\rm s}}/(n_eB_\parallel)$ and $\eta_l=(l_{\rm s}/\Delta R)^{1/2}$.
According to Eq.(\ref{sigma}) and Eq.(\ref{RM}), the specific luminosity of synchrotron radiation is 
\be
L_{\nu}&=& \frac{64\pi^3}{27}\frac{\zeta_e\gamma_{\rm th}^2}{\xi_{nB}\eta_l}m_ec^2R^2\sigma_{\rm RM} \nonumber \\
& \simeq& 5.7\times10^{29}~{\rm erg~s^{-1}~Hz^{-1}}
\left(\frac{\xi_{nB}}{0.1}\right)^{-1}\left(\frac{\eta_l}{0.1}\right)^{-1} \nonumber \\
&\times&\left(\frac{\zeta_e\gamma_{\rm th}^2}{0.01}\right) \left(\frac{\sigma_{\rm RM}}{10~{\rm rad~m^{-2}}}\right)\left(\frac{R}{1~{\rm pc}}\right)^2. \label{lum1}
\ee
The results of Eq.(\ref{lum}) and Eq.(\ref{lum1}) suggest that FRBs with large RM and $\sigma_{\rm RM}$ values tend to be associated with compact PRS. This is consistent with the observations of FRB 121102 and FRB 190520B \citep{Yang20a,Feng22}. 

Another way to estimate the radio luminosity is to assume that electrons radiating in the GHz band contribute a fraction of the total pressure, so that it may scales with the magnetic pressure, i.e. $\gamma^2(dn_e/d\gamma)m_ec^2\sim B^2/8\pi$. Because the electron distribution satisfies $n_e(\gamma)\propto\gamma^{-2}$ in most astrophysical scenarios, $\gamma^2(dn_e/d\gamma)$ is roughly independent of $\gamma$. 
According to Eq.(\ref{gammaghz}), the specific luminosity is given by 
\be
L_\nu &\simeq& \gamma \frac{dn_e}{d\gamma} \left(\frac{4\pi}{3}R^3\right) P_\nu\simeq3.7\times10^{27}~{\rm erg~s^{-1}Hz^{-1}} \nonumber\\
&\times&\left(\frac{\nu}{1~{\rm GHz}}\right)^{-1/2}\left(\frac{B}{1~{\rm mG}}\right)^{7/2}\left(\frac{R}{1~{\rm pc}}\right)^{3}
\ee
If we further assume that the magnetic energy in the magnetized plasma screen is contributed by the central neutron star engine with a magnetic energy $E_B$, i.e. $(B^2/8\pi)(4\pi R^3/3)\sim \beta_B E_B$ with $\beta_B<1$, then the above equation could be written as
\be
L_\nu &\simeq&2.4\times10^{28}~{\rm erg~s^{-1}Hz^{-1}} \eta_B\nonumber\\
&\times&\left(\frac{\nu}{1~{\rm GHz}}\right)^{-1/2}\left(\frac{B}{10~{\rm mG}}\right)^{3/2}\left(\frac{E_B}{10^{48}~{\rm erg}}\right)
\ee
This shows that the PRS should only be detected from sources with a strongly magnetized environment, e.g. $B\gtrsim 10~{\rm mG}$. As the nebula expands with time, the magnetic energy drops due to adiabatic losses, and the magnetic field strength also decreases due to increasing volume, thus one may only expect to detect bright PRS from very young systems.

\section{Conclusions and Discussions}\label{sec4}

Recently, some newly discovered repeating FRBs were found to possess complex polarization properties and be associated with compact PRS. For example, FRB 121102 showed a significant RM evolution during a long term \citep{Michilli18,Hilmarsson21}, meanwhile, it was associated with a compact PRS \citep{Chatterjee17}. FRB 180301 exhibited significant PA swings on a timescale of $\sim 10~{\rm ms}$ while maintaining a large polarization degree \citep{Luo20b}. FRB 190520B also was found to be associated with a compact PRS and have an extremely significant DM contribution by its host galaxy \citep{Niu21}. FRB 20201124A showed significant, irregular, short-time variation of the Faraday rotation, and the frequency spectra of the polarized components of some bursts appear to show clear oscillating structures \citep{Xu21}. 
Very recently, \citet{Feng22} reported that active repeaters exhibit conspicuous frequency-dependent depolarization and a strong correlation between $\sigma_{\rm RM}$ and $\tau_{\rm s}$ ($\sigma_{\rm RM}\propto\tau_{\rm s}^{1.0\pm0.2}$), meanwhile, the FRBs with compact PRS tend to have extreme RM scatter. 
The observational properties imply that these FRB sources are located in magnetized plasma environments, likely a supernova remnant or a pulsar wind nebula.

Theories on temporal scattering and depolarization of FRBs have been discussed on some previous works \citep[e.g.,][]{Xu16,Beniamini20,Beniamini21}.
In this work, we mainly focus on some theoretical predictions on the relations among temporal scattering, depolarization by RM scatter, and persistent radio emission contributed by the magnetized plasma environment close to a repeating FRB source.
First, we predict a relation between RM scatter and temporal scatting time, as shown by Eq.(\ref{sigmatau1a}) and Eq.(\ref{sigmatau1b}). Since $B_\parallel\propto\Delta{\rm RM}/\Delta{\rm DM}$ is involved \citep{Katz18,Katz21}, such a relation could be tested once the root-mean-squared variations of DM and RM are measured for repeating FRBs in the future. 
Furthermore, if one assumes that the turbulent plasma satisfies energy equipartition with $B\propto n_e^{1/2}$, then the relation between RM scatter and temporal scatting time would become Eq.(\ref{sigmatau2a}) and Eq.(\ref{sigmatau2b}), which gives $\sigma_{\rm RM}\propto\tau_{\rm s,0}^{(0.54-0.67)}$ for the Kolmogorov turbulence. This is shallower than the observed relation by \citet{Feng22}. 
If the turbulent plasma satisfies magnetic frozen conditon with $B\propto n_e^{2/3}$, the predicted $\sigma_{\rm RM}$-$\tau_{\rm s}$ becomes Eq.(\ref{sigmatau4a}) and Eq.(\ref{sigmatau4b}), which gives $\sigma_{\rm RM}\propto\tau_{\rm s}^{(0.58-0.72)}$ for the Kolmogorov turbulence. 
If the turbulent plasma is shock compressed with $B\propto n_e$, the predicted $\sigma_{\rm RM}$-$\tau_{\rm s}$ becomes Eq.(\ref{sigmatau3a}) and Eq.(\ref{sigmatau3b}), which gives $\sigma_{\rm RM}\propto\tau_{\rm s}^{(0.65-0.83)}$ for the Kolmogorov turbulence. This is closer to the observed relation \citep{Feng22}. 
Besides, since the RM scatters measured by the frequency-dependent depolarization are much less than the absolute values of RMs of most repeaters \citep{Feng22}, it implies that a large-scale magnetic field may exist in the plasma screen, or the observed RM is from a different region.
Very recently, \citet{Anna-Thomas22} reported that the RM of FRB 190520B is rapidly varying with a large amplitude. Since the $\sigma_{\rm RM}$ affecting the depolarization is mainly contributed by the small-scale fluctuation $\delta(n_eB_\parallel)$ with scale much less than $l_{\rm s}$ given by Eq.(\ref{lsca}), the result of $|{\rm RM}|\gg\sigma_{\rm RM}$ implies that the observed varying RM is mainly contributed by the large-scale magnetic field and $\delta B\ll B$ is required.

We then discuss the relation between RM and the luminosity of the PRS for repeating FRBs. Different from most previous works with PRS depending on some specific astrophysical scenarios \citep[e.g.,][]{Yang16,Dai17,Margalit18}, in this work we make a general discussion only assuming that PRS and RM originate from the same region and PRS is produced by synchrotron radiation. We find that the thermal component of accelerated electrons could not be ultra-relativistic.
We also predict that the larger the RM and/or RM scatter, the brighter the persistent radio emission from the plasma screen. This result is consistent with the observation of FRB 121102 \citep{Chatterjee17,Michilli18,Yang20a}. According to this picture, FRB 190520B with a compact PRS should also have a relativelty large RM value. This is consistent with the preliminary analysis of \citep{Niu21} and the very recent observational result \citep{Anna-Thomas22}. On the other hand, under the assumption that the plasma screen's magnetic energy originates from the activities in the magnetosphere of a neutron star, the brightness of the PRS will fade with time as the nebula expands. Thus, FRB sources with compact PRSs might be very young as proposed in some previous works \citep[e.g.,][]{Margalit18,Zhao21b}.

\acknowledgments
We thank the anonymous referee for helpful comments and suggestions.
We also thank Siyao Xu for the constructive discussion about the MHD turbulence, and thank Jonathan Katz, Kohta Murase, Yuan-Hong Qu, Fa-Yin Wang, Zhao-Yang Xia for helpful discussions.
This work has been supported by National Natural Science Foundation of China grant No. 11988101 and No. 12003028, and China Manned Spaced Project (CMS-CSST-2021-B11). WL is supported by Lyman Spitzer, Jr Fellowship at Princeton University. Y.F. is supported by Key Research Project of Zhejiang Lab No. 2021PE0AC03.

\end{document}